\documentclass[aps,prb,reprint,superscriptaddress,amsmath,amssymb]{revtex4-2}
\usepackage[utf8]{inputenc} 

\DeclareUnicodeCharacter{0307}{} 
\usepackage{xcolor}
\usepackage{graphicx}
\DeclareGraphicsExtensions{.png,.pdf,.tif,.jpeg}
\DeclareUnicodeCharacter{03BA}{\ensuremath{\kappa}}

\usepackage{parskip}
\usepackage{siunitx}
\usepackage{amsmath}
\begin{document}


\title{Conventional Type-II Superconductivity in 2H-TaSeS}

\author{K. Yadav}
\author{M. Lamba}
\author{M. Singh}
\author{A. Kumar}
\author{S. Patnaik}
\email{spatnaik@jnu.ac.in}
\affiliation{School of Physical Sciences, Jawaharlal Nehru University, New Delhi-110067, India}


\date{\today}
\begin{abstract}
Superconductors based on transition metal dichalcogenides are of substantial current relevance towards attaining topological superconductivity. Here we report a detailed study on synthesis and electromagnetic characterization of high-quality single crystals of TaSeS. A superconducting transition is confirmed at $4.15K$ with coexisting charge density wave onset at $66K$. The temperature dependence of RF penetration depth indicates s-wave characteristics in the weak coupling limit. A moderate electronic anisotropy is observed in upper critical fields with a value of $1.52$. DFT calculations confirm the possibility of superconducting behavior of TaSeS and also suggest that the most stable structure belongs to $P6_3mc$ space group. Negative values in phonon dispersion curves verify the possibility of co-existing CDW in 2H-TaSeS. Flux pinning disappears at $H^{*}_{\bot ab(0)}\sim0.65H_{c2,\bot ab}(0)$ and $H^{*}_{\parallel ab}(0)\sim0.74H_{c2,\parallel  ab}(0)$. Arrhenius plots show power law dependence of activation energy with respect to magnetic field. Overall all characteristics imply TaSeS to be a classic Type-II superconductor without any evidence for topological superconductivity.
\begin{description}
\item[PACS numbers] 74.25.Fy, 74.25.Ha, 74.25.Qt, 74.50.+r
\item[Keywords]
charge density waves, anisotropy, flux flow, flux pinning, penetration depth
\end{description}
\end{abstract}

\maketitle

\begin{figure*}
\includegraphics[width=1\textwidth,height=6cm]{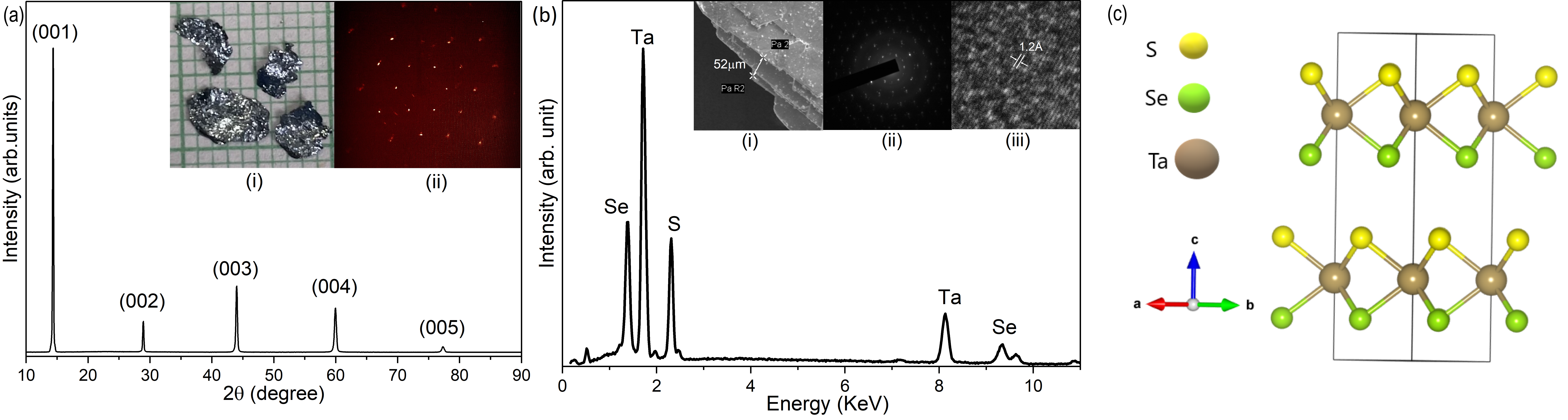} 
  \caption{(a) X-ray diffraction pattern of single crystal of TaSeS and inset (i) shows close image of single crystals and inset (ii) shows the single crystal diffraction pattern of the single crystal. (b) EDAX data of TaSeS and inset (i) shows the SEM image that confirms the layered morphology, inset (ii)  shows the SAED pattern of TaSeS and the HRTEM images show $[00l]$ planes. (c) Unit cell of 2H-TaSeS.}
\end{figure*}
\section{\label{sec:level1}INTRODUCTION}
The layered transition metal dichalcogenides (TMD), with van der Waals bonding between their two-dimensional constituents, offer an exceptional platform to study the vagaries of electronic correlations in quantum condensed matter. Of great current interest is the tantalum based dichalcogenides that promises the possibility of topological superconductivity \cite{sato2017topological,leijnse2012introduction}. In specific, both 2H-TaSe$_2$ and 2H-TaS$_2$ show superconductivity in the milli-Kelvin range with coexisting charge density wave (CDW).  Moreover, a time reversal symmetry broken superconducting state is recently reported in 4H$_b$-TaS$_2$, a phase which is constituted by alternating layers of 1T-TaS$_2$ (gapless spin liquid ground state) and 1H-TaS$_2$ (two dimensional superconductor with CDW). In general TMDs display a variety of characteristics, including superconductivity, charge density wave order (CDW) \cite{li2013influence,sugawara2016unconventional,nakata2021robust}, electron-electron correlation \cite{nakata2021robust}, spin-orbit coupling \cite{lutchyn2010majorana}, and topological properties \cite{hsu2017topological,xu2014topological,hooda2020electronic}. The combination of CDW phase and topological properties make TMDs an ideal system to investigate the interaction between superconductivity and topology of the electronic band structure.\par
Depending on the composition, TMDs exhibit different properties and can be superconductors (e.g., TaS$_2$, NbSe$_2$), metals (e.g., VSe$_2$, NbS$_2$), semimetals (e.g., WTe$_2$, TiSe$_2$), semiconductors (e.g., MoS$_2$, WS$_2$), and insulators (e.g., MoSe$_2$, WSe$_2$) \cite{lv2015transition}. Strong spin-orbit coupling, proximity-induced superconductivity, and the presence of magnetic or spin-polarized adatoms on the surface \cite{kormanyos2015landau,lee2015ultimately,saxena2000superconductivity} of TMDs all interact to cause the appearance of possible topological superconductivity in such materials. These characteristics result in nontrivial topology in momentum space for unconventional superconducting states, which lead to the development of Majorana zero modes near the material boundaries or defects. One well-known compound exhibiting topological superconductivity is the Molybdenum Disulfide (MoS$_2$). Majorana zero modes can be seen at the borders of a MoS$_2$ monolayer device connected to a superconducting electrode \cite{mourik2012signatures}. TMDs have the potential to serve as a platform for topological quantum computation, as evidenced by the detection of zero-bias conductance peaks, which hints at Majorana fermions. Exciting characteristics, such as proximity-induced superconducting state and appearance of Andreev bound states, have also been discovered through the study of WS$_2$-based heterostructures \cite{lian2020anisotropic}.\par
In addition to being superconductors, TaS$_2$ and TaSe$_2$ display a variety of intriguing electronic phenomena. As an illustration, TaS$_2$ shows a charge density wave (CDW) transition at low temperatures, which results in the emergence of a periodic lattice distortion and a Peierls gap \cite{li2013influence,ritschel2015orbital}. It has been discovered that this CDW phase coexists with superconductivity, creating a rich interplay between the two phenomena. Additionally, the interaction of spin-orbit coupling with electronic correlations in TMDs can result in topological phases like Weyl semimetals and topological insulators, substantially enhancing the range of topological events in TaS$_2$ and TaSe$_2$ \cite{woolley1977band}. Similarly, recent works on TaSeS show superconductivity at
3.9K as a result of CDW suppression \cite{patra2022two}. Also, it is reported to be an unconventional superconductor with Fulde-Ferrell-Larkin- Ovchinnikov phase in large magnetic fields \cite{patra2022two}.\par
In the present work, we report synthesis of high quality single crystals of 2H-TaSeS. Detailed magneto-transport and magnetization measurements show anisotropic electronic behavior of single crystals. Utilizing the transport data of the compound under different magnetic fields, a thorough examination of activation energy and flux flow mechanisms has been conducted. The flux depinning is described by the irreversible fields which indicates that there is a weak vortex fluctuation inside the compound. The order parameter symmetry is found to be s-wave in contradiction to the expected p$_x$+ip$_y$ order parameter for possible topological superconductor \cite{neha2019time}. We find that TaSeS adheres well to the prediction of Type-II BCS superconductivity.
\section{EXPERIMENTAL TECHNIQUES}
Chemical vapor transport (CVT) technique was used for the synthesis of single crystals of 2H-TaSeS. For that purpose, the stoichiometric ratios of Tantalum (Alpha Aesar, $99.9\%$), Selenium (Aldrich, $99.99\%$), and Sulfur (Aldrich, $99.98\%$) were taken and thoroughly grounded for 30 minutes and molded into pellets with the help of hydraulic press. Pellets were then put into vacuum sealed quartz tube with iodine as transport agent. The tube was then placed in a two-zone furnace at a temperature gradient of $100^\circ C$ where temperatures of hot and cold zone were $850^\circ C$ and $750^\circ C$ respectively. After 10 days, shiny single crystals of TaSeS were obtained in the low temperature zone of the tube. Room temperature X-Ray diffraction (XRD) method was used to confirm the orientation of crystal plane using Rigaku Miniflex-600 X-Ray diffractometer with Cu-K$\alpha$ radiation $(1.54056\si{\angstrom})$. Laue diffraction pattern was obtained with the help of Bruker X-Ray diffractometer with Mo-K$\alpha$ radiation $(0.71\si{\angstrom})$. Images from scanning electron microscopy (SEM) and energy dispersive x-ray spectroscopy (EDAX) were obtained using a Bruker AXS microanalyzer and a Zeiss EVO40 SEM analyzer, respectively. A JEOL transmission electron microscope was used to perform high resolution transmission electron microscopy (HRTEM) measurements. Magnetic penetration depth measurements using tunnel diode oscillator (TDO) technique and magnetotransport measurements  were carried out with a Cryogen Free Magnet (Cryogenic, $8T$) and temperature dependent magnetization measurements were done using Vibrating Sample Magnetometer (VSM) probe of a Physical Property Measurement System (Cryogenic, $14T$). Vienna Ab-initio Simulation Package (VASP) and Wien2K codes were used for Density Functional Theory (DFT) calculations.
\subsection{\label{app:subsec}Crystal Structure analysis}
  Fig. 1(a) shows the X-ray diffraction pattern of crystal flake of TaSeS. All diffraction peaks strongly correspond to $[00l]$ plane that confirms the single crystalline nature of the sample. Single crystals of 2H-TaSeS of millimeters sizes are shown in the inset (i) of the Fig. 1(a). Typical large area crystal flakes are of dimensions  $6mm\times4mm\times0.02mm$. Single crystalline nature of the sample was further confirmed by the Laue spots as shown in the inset (ii) of Fig. 1(a). The EDAX pattern of TaSeS is shown in Fig. 1(b). The elemental composition was obtained through quantitative study of the EDAX data from several points on the sample surface. The average atomic percentage of Ta, Se and S are observed to be $37\%$, $27\%$ and $36\%$ respectively. These values are slightly away from the stoichiometric composition attempted during synthesis. Furthermore, the evidence of off-stoichiometry compound can be seen in the XRD data. Here absence of extra peaks corresponding to other planes shows that there is a definite arrangement of atoms in the unit cell instead of random substitutions. Layered morphology of the sample was clearly shown in scanning electron microscope (SEM) image (inset (i) of Fig. 1(b)). The  Selected Area Electron Diffraction (SAED) pattern of TaSeS can be seen in the inset (ii) of Fig. 1(b). This depicts a diffused and azimuthal spread of the superlattice satellite spots connected to each TaSeS Bragg points and this azimuthal spread is caused by the large atomic radii of Ta. The similar diffusion pattern was earlier reported in layered dichalogenides and it was found that the potential for a charge density wave transition at lower temperature is foreshadowed by the establishment of a superlattice at a certain higher temperature \cite{wilson1975charge}. Inset (iii) of Fig. 1(b) shows the HRTEM images of TaSeS along $[00l]$ plane which provides the crystalline planes with d-spacing of $1.2\si{\angstrom}$. Fig. 1(c) shows the schematic structure of the unit cell of TaSeS with $P6_3mc$ space group which consists of coupled sandwich layers S–Ta–Se with strong covalent bonding. Weak van der Waals interaction holds the adjacent sandwiched layers to form the bulk crystal. Using Fullprof software XRD data was refined to calculate the Lattice parameters that come out as $a = b = 3.37\si{\angstrom}$ and $c = 12.34\si{\angstrom}$ which is almost similar to the reported data \cite{patra2022two}.
  \begin{figure}
\includegraphics[width=0.45\textwidth,height=6cm]{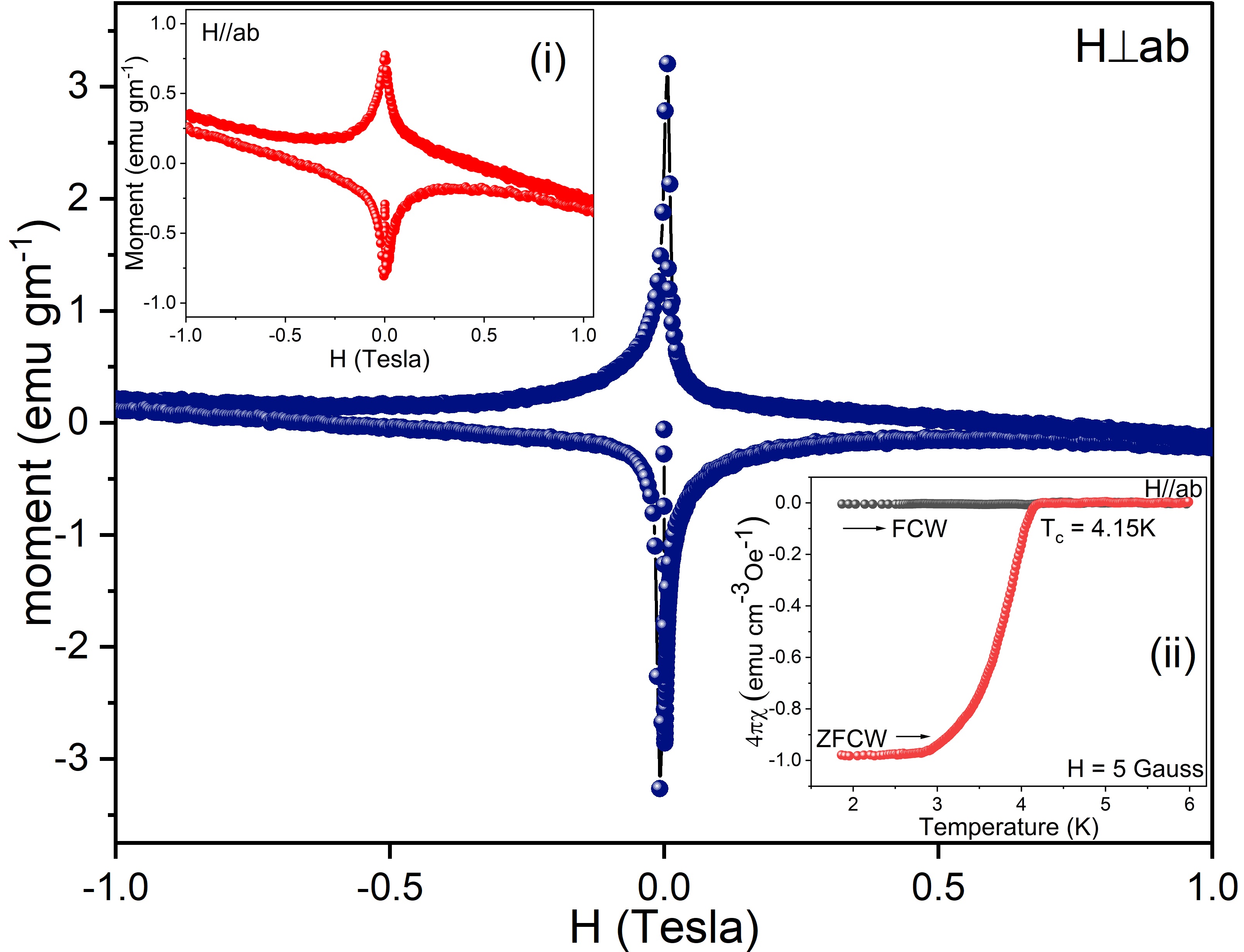} 
  \caption{(M-H loop plotted for $H\bot ab$ orientation. Inset (i) shows the M-H loop and inset (ii) shows dc magnetization (ZFC and FC) along in $H\parallel ab$ orientation.}
\end{figure}
  \subsection{\label{app:subsec}Magnetization}
  Fig. 2 and its inset (i) show the magnetization Vs. magnetic field (M-H) loops were taken in both the orientations $(H\bot ab)$ and $(H\parallel ab)$ respectively. Lower critical field $H_{c1}$ was calculated using the Ginzburg-Landau equation given by $H_{c1}(T) = H'_{c1}(0) [1-(T/T_c)^2]$ \cite{rani2018temperature}. The value of $H'_{c1,\bot ab}(0)$ comes out to be $2.4 mT$ but there is substantial demagnetization factor in this orientation. Therefore, $H_{c1,\bot ab}(0)$ is calculated by the formula, $H_{c1,\bot ab}(0) = H'_{c1,\bot ab}(0)/\sqrt{tanh(0.67\times\frac{c}{a})}$, where $c$ and $a$ are the thickness of sample and length of sample respectively. This yields $H_{c1,\bot ab}(0)$ value to be $30.8 mT$. Since demagnetization effect is negligible in $H\parallel ab$ orientation, the value of $H_{c1,\parallel ab}(0)$ is directly calculated out to be $23.6 mT$. DC magnetization measurement was done on the single crystal in zero field cooled warming (ZFCW) and field cooled warming (FCW) mode with an applied $5 Gauss$ field in $H\parallel ab$ orientation as shown in the inset (ii) of Fig. 2. This shows the superconducting transition. The transition temperature is determined to be $T_c = 4.15K$ which is among highest transition temperature reported in this system TaSe$_{1-x}$S$_x$. From ZFC susceptibility, the superconducting shielding fraction came out to be $98\%$.
  \subsection{\label{app:subsec}Resistivity}
  Fig. 3(a) shows remperature dependent resistivity of TaSeS single crystal with linear four probe method at a dc current of $2mA$ in the absence of any external field. The inset (i) of Fig. 3(a) confirms the resistive superconducting transition at $4.15K$ which is equal to the transition temperature measured in magnetization measurements. The transition from normal to superconducting state is quite sharp with a width of $\Delta T_c(=T_{c_{onset}} - T_{c_{zero}}) = 0.14K$. This indicates good grains connectivity which reflects the quality of crystals.
   \begin{figure*}
  \includegraphics[width=0.9\textwidth,height=12cm]{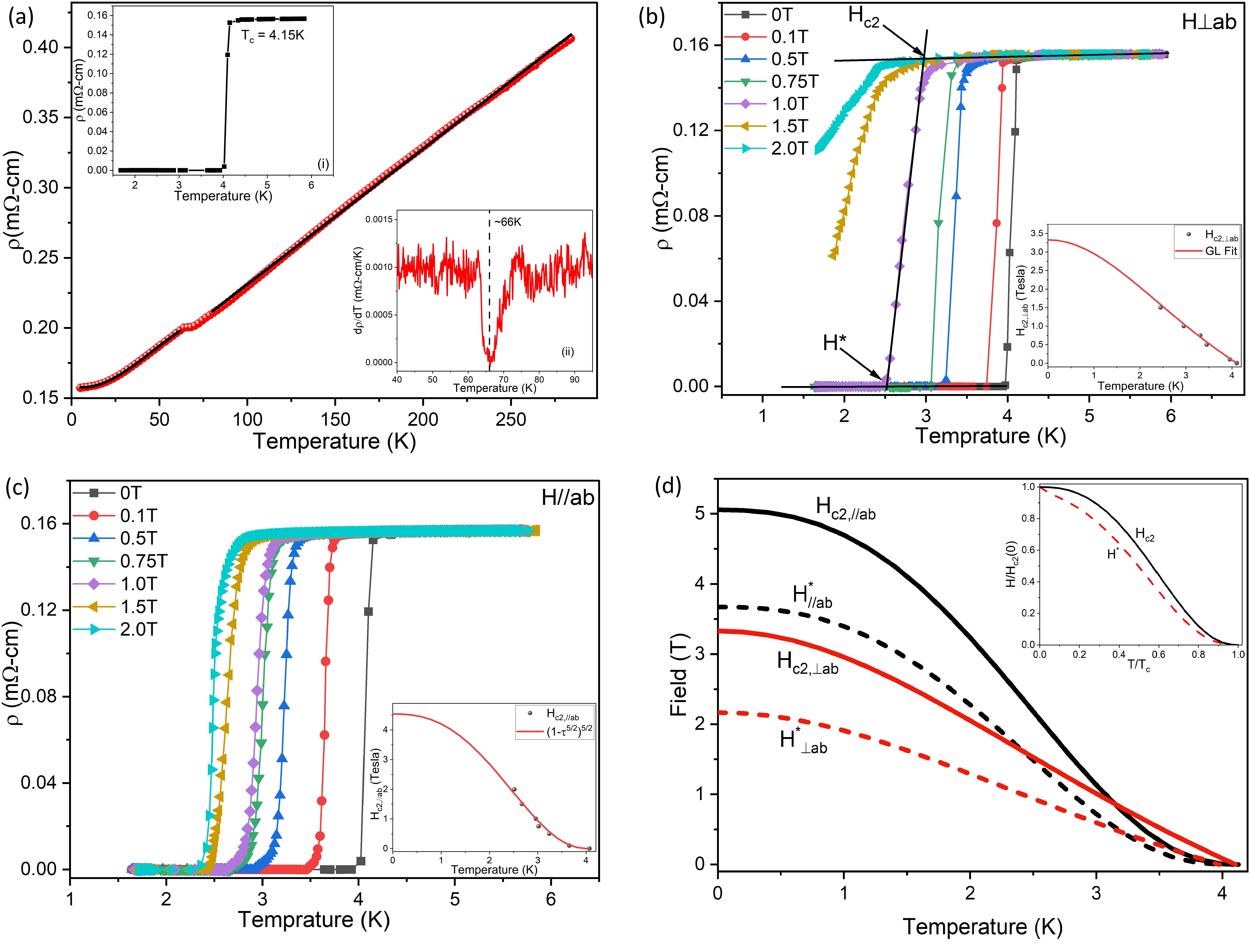} 
  \caption{(a) Zero field resistivity data above transition temperature and inset (i) shows the superconducting state at $4.15K$ and inset (ii) shows the resistivity derivative data with CDW transition at $66K$. (b) Resistivity v/s temperature data for different fields $(H = 0T, 0.1T, 0.5T, 0.75T, 1.0T, 1.5T, 2.0T, 2.5T)$  in $H\bot ab$ orientation and inset shows the GL fitting of upper critical field $H_{c2}(T)$ for $H\bot ab$ orientation. (c) Resistivity v/s temperature data for different fields $(H = 0T, 0.1T, 0.5T, 0.75T, 1.0T, 1.5T, 2.0T, 2.5T)$ in $H\parallel ab$ orientation and inset shows the upper critical field $H_{c2}(T)$ fit for $H\parallel ab$ orientation. (d) Fitted upper critical field $H_{c2}$ and irreversibility field $H^*$ calculated by resistivity measurements and inset shows the thermal depinning line for $\alpha = 0.15$ in reduced temperature and fields coordinates.}
\end{figure*}
  The residual resistivity ratio (RRR= $\rho$($290K$)/$\rho$($5K$)) of the sample is about $2.7$ and this sample shows metallic behavior above transition temperature. Temperature dependent resistivity data above $80K$ shows a linear behavior with slope $9.66\times10^{-4}m\Omega-cmK^{-1}$. This linear behavior is same as the behavior exhibited by the metals due to the scattering of electrons by acoustic phonons \cite{li2013influence}. For low temperatures there comes the Bloch-Grüneisen model \cite{mamedov2007new} which states that most of energetic phonons would have momentum smaller than $\hbar k_F$ which is the momentum of electrons at Fermi surface. This also implies that the conducting electrons only scatters at small angles when they emit or absorb phonons. Bloch-Grüneisen model typically shows $\rho \sim T^5$ dependency. But a special case called Bloch–Wilson limit \cite{imai2012superconductivity,caton1982analysis}, shows $\rho \sim T^3$ dependence at low temperature which fits better with our data. This $T^3$ dependency is assigned to s-d inter-band scattering, which is common among transition metals. At about $66K$ there was a humped in the resistivity Vs. temperature curve. This was also confirmed by the resistivity derivative with respect to temperature as shown in inset (ii) of Fig.3(a). This is a clear evidence of charge density wave (CDW) that is similar to its parent sample TaS$_2$ \cite{ge2010charge} where superconducting transition temperature and CDW transition occur at $1K$ and $77K$ respectively.\par
  Magneto-resistance was measured in presence of different magnetic fields in both $H\parallel ab$ and $H\bot ab$ directions. When magnetic field was applied to the crystal for $H\bot ab$ orientation, the superconducting transition temperature shifts to the lower values as shown in Fig.3(b). In $H\bot ab$ orientation, $H_{c2,\bot ab}$ Vs. temperature curve was fitted using the Ginzburg-Landau equation i.e., $H_{c2,\bot ab}(T)$ = $H_{c2,\bot ab}(0)[(1- \tau^2)/(1+ \tau^2)]$, where $\tau = T/T_c$ \cite{xing2016anisotropic}. As shownn in the inset of Fig. 3(b) the $H_{c2}(T)$ phase diagram fits well with the Ginzburg-Landau equation. The estimated value of $H_{c2,\bot ab}(0)$ is $3.3T$. Fig. 3(c) shows the magneto-resistance data for $H\parallel ab$ orientation. $H_{c2,\parallel ab}$-temperature curve could not be fitted using the Ginzburg-Landau equation as there is non-linearity in high temperature range. such departure of linear behaviour behavior in $H_{c2}$-temperature curve was previously reported in TMDs and has been assigned to multiband effects and dimensional crossover \cite{hohenberg1967anisotropy,klemm1975theory}. The best fit for this curve was obtained by fitting the equation $H_{c2,\parallel ab}(T)$ = $H_{c2,\parallel ab}(0)(1- \tau^{5/2})^{5/2}$ \cite{singh2022superconductivity} as shown in inset of Fig. 3(c). The extrapolated upper critical field $H_{c2,\parallel ab}(0)$ is estimated to be $5.05T$. The anisotropy in upper critical fields $[\gamma = H_{c2,\parallel ab}(0)/H_{c2,\bot ab}(0)]$ comes out to be $1.52$.
  \subsection{\label{app:subsec}Characteristic lengths}
   The upper critical field was used to calculate the Ginzburg-Landau coherence lengths for both orientations ($\xi\parallel ab$ and $\xi\bot ab$) using the formulas
   \begin{equation}
   H_{c2,\bot ab}(0) = \frac{\phi_{0}}{2\pi\xi^{2}_{\parallel ab}(0)}
   \end{equation}
   and  \begin{equation}
   H_{c2,\parallel ab}(0) = \frac{\phi_{0}}{2\pi\xi_{\parallel ab}(0) \xi_{\bot ab}(0)}
   \end{equation} 
   where $\phi_0$ is the flux quantum and has the value equal to $2.07\times10^{-15}Wb$. The estimated values of coherence lengths were $\xi_{\parallel ab}(0)$ = $9.95 nm$ and $\xi_{\bot ab}(0)$ = $6.56 nm$. For penetration depth, Ginzburg-Landau parameters $\kappa_{\parallel ab}(0)$ and $\kappa_{\bot ab κ}(0)$ were calculated using the equations
   \begin{equation}
   \kappa_{\parallel ab}(0) = \sqrt{\frac{\lambda_{\parallel ab} \lambda_{\bot ab}}{\xi_{\parallel ab}\xi_{\bot ab}}}
   \end{equation}
   and 
   \begin{equation}
   \kappa_{\bot ab}(0) = \frac{\lambda_{\parallel ab}}{\lambda_{\bot ab}}
   \end{equation}
   The value of $\kappa$ can be obtained by the formula $H_{c2}(0)/H_{c1}(0) = 2\kappa^2/\ln\kappa$ \cite{singh2022superconductivity}. Thus, the value of $\kappa_{\bot ab}$ was calculated by $H_{c2,\bot ab}(0)/H_{c1,\bot ab}(0)$ = $2\kappa^2_{\bot ab}/\ln \kappa_{\bot ab}$ and that came out to be $11.4$.Similarly, the value of $\kappa_{\parallel ab}$ was $17.5$ which is greater than $1/\sqrt{2}$. This indicates the type-II behavior of superconducting phase of TaSeS \cite{hake1967upper}. These numbers are used to calculate the values of penetration depth in both orientations which are estimated to be $\lambda_{\parallel ab}(0) = 114 nm$ and  $\lambda_{\bot ab}(0) = 174 nm$.
 \subsection{\label{app:subsec}Thermal depinning of vortex lattice}
 To determine the irreversibility line, we suppose that the thermal depinning occurs at $H^*(T)$ known as thermal depinning field, which is the point where the mean-squared displacements of the vortex lines become equal to square of coherence length i.e., $u^2(T, H^*) = \xi^2(T)$ \cite{patnaik2001electronic}. The irreversible field $H^*(T) = b(t)H_{c2}(T)$, where $b(t)$=$H/H_{c2}$ is the reduced magnetic field. Fig. 3(d) shows the resistivity determined $H_{c2}(T)$ and $H^*(T)$ lines fitted for both orientations. The graph indicates that values of $H_{c2}(T)$ and $H^*(T)$ are higher in the orientation where magnetic field is parallel to ab plane. Values of irreversiblity fields $H^*_{\bot ab}(0)$ and $H^*_{\parallel ab}(0)$ were calculated by the fitted graph \cite{fuchs2001upper,matsushita1990flux} and that come out to be $2.16T$ and $3.76T$ respectively. Also, $H_{c2}(T)$ and $H^*(T)$ are separated with a significant value with $H^*_{\bot ab}(0)\sim0.65 H_{c2,\bot ab}(0)$ and $H^*_{\parallel ab}(0)\sim0.74 H_{c2,\parallel ab}(0)$. The value of $b(t)$ can be calculated from the equation $t^2 = g(b)/[\alpha^2+g(b)]$, where $\alpha$ is the vortex thermal fluctuation strength and $g(b) = b(t)[1-b(t)]^3\ln[2+{2b(t)}^{-1/2}]$. The value of $\alpha$ can be calculated using the formula
 \begin{equation}
     \alpha = \frac{4\sqrt{2} \mu_0\pi\gamma\kappa^2\xi(0)k_BT_c)}{\phi_0^2}
 \end{equation} 
 and thus values of $\alpha_{\parallel ab}$ and $\alpha_{\bot ab}$ came out to be $0.059$ and $0.016$. These small values of $\alpha$ indicate the weak vortex fluctuations. Inset of Fig. 3 (d) shows the calculated depinning line $H^*(T)$ for $\alpha = 0.059$ compared with the upper critical field in reduced field and temperature coordinates.
 \begin{figure}
\includegraphics[width=0.45\textwidth,height=6cm]{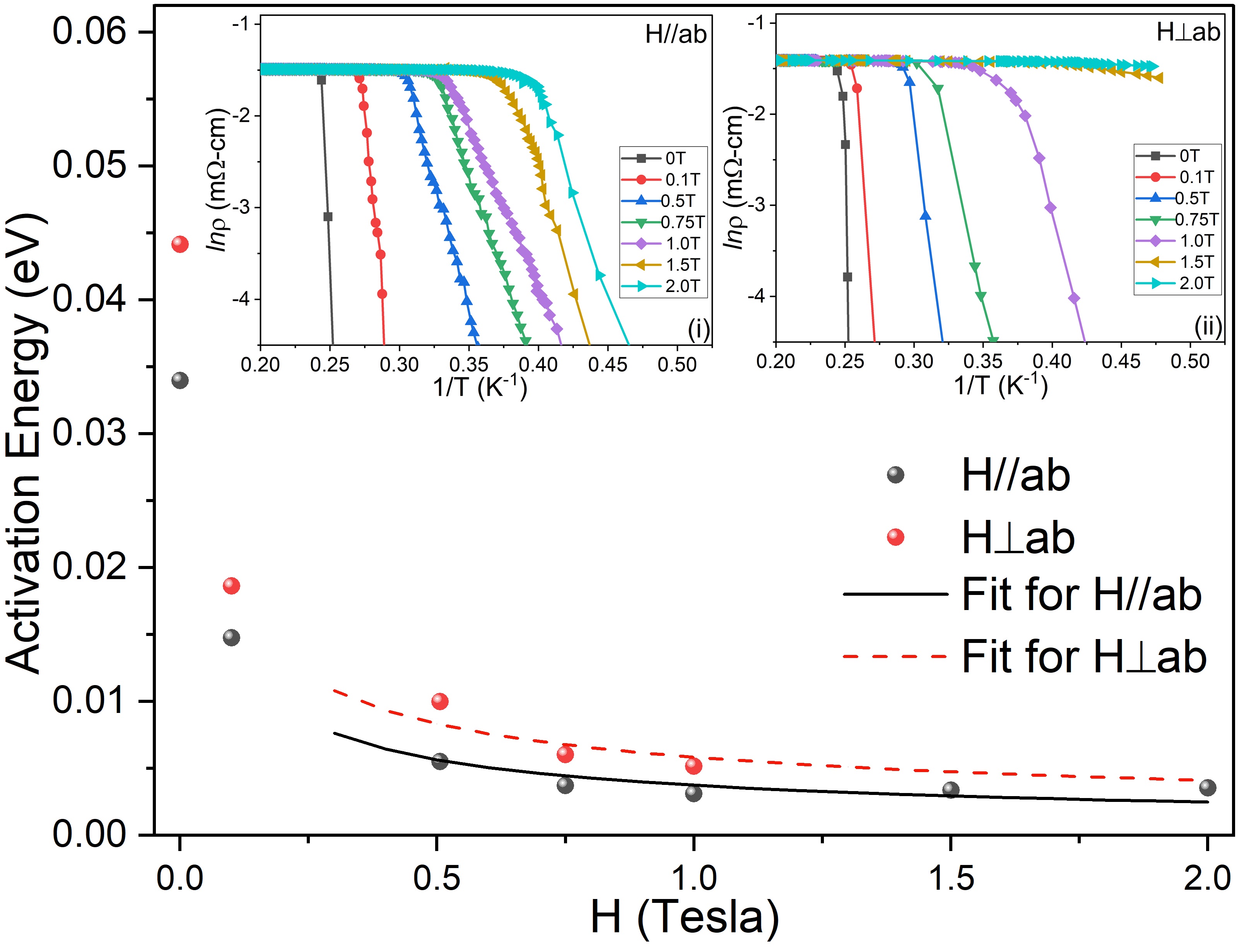} 
  \caption{Activation energy behavior in magnetic field above for both orientations $H\bot ab$ and $H\parallel ab$. Insets (i) and (ii) shows Arrhenius plots $(\ln\rho$ v/s $1/T)$ for $H\parallel ab$ and $H\bot ab$ respectively.}
  \end{figure}
 \subsection{\label{app:subsec}Activation energy}
 There is a broadening in the transition in both orientations, which increase as the external field is increased. This broadening in magneto-resistance is due to the thermal activated flux flow (TAFF) of the vortices. Arrhenius law \cite{akyol2015effect,sudesh2012effect} was used to find the activation energy of the TAFF
 \begin{equation}
     \rho = \rho_0exp(-U_0/k_BT)
 \end{equation}
 where $\rho_0$ is resistivity at normal state, $U_0$ is activation energy of flux flow and $k_B$ is Boltzmann constant. Insets (i) and (ii) of Fig. 4 show the Arrhenius plots of the resistivity $(\ln\rho$ Vs. $1/T)$ with external magnetic fields for both $H\parallel ab$ and $H\bot ab$ orientations respectively. The transition region shows the region for thermally activated flux flow and slope from the linear fit gives the value of activation energy $U_0$. Fig. 4 shows the graph between activation energy and applied magnetic field. At magnetic fields below $\sim 0.4 T$, large value of pinning energy illustrates strong vortex pinning at low fields. This shows that lower magnetic fields below $\sim 0.4 T$ can only penetrate into the intergranular regime.  As shown in Fig. 4 data fits well with the power law equation $U\propto H^{-\zeta(H)}$, where $\zeta$ is the function of applied magnetic field. It can be clearly observed that for both of the orientation, activation energy obeys the power law behavior with magnetic field, but has a variance in exponents which is possibly linked with the microstructure of the crystal. The value of $\zeta$ depends on the range and the orientation of the applied magnetic field. The value of $\zeta\parallel ab$ and $\zeta\bot ab$ come out to be $0.59$ and $0.51$ respectively for external field above $\sim 0.4 T$. The small value of $\zeta$ in higher fields and field dependency of activation energy implies that in both orientations TaSeS shows both flux flow and flux pinning.
 \begin{figure}
\includegraphics[width=0.45\textwidth,height=6cm]{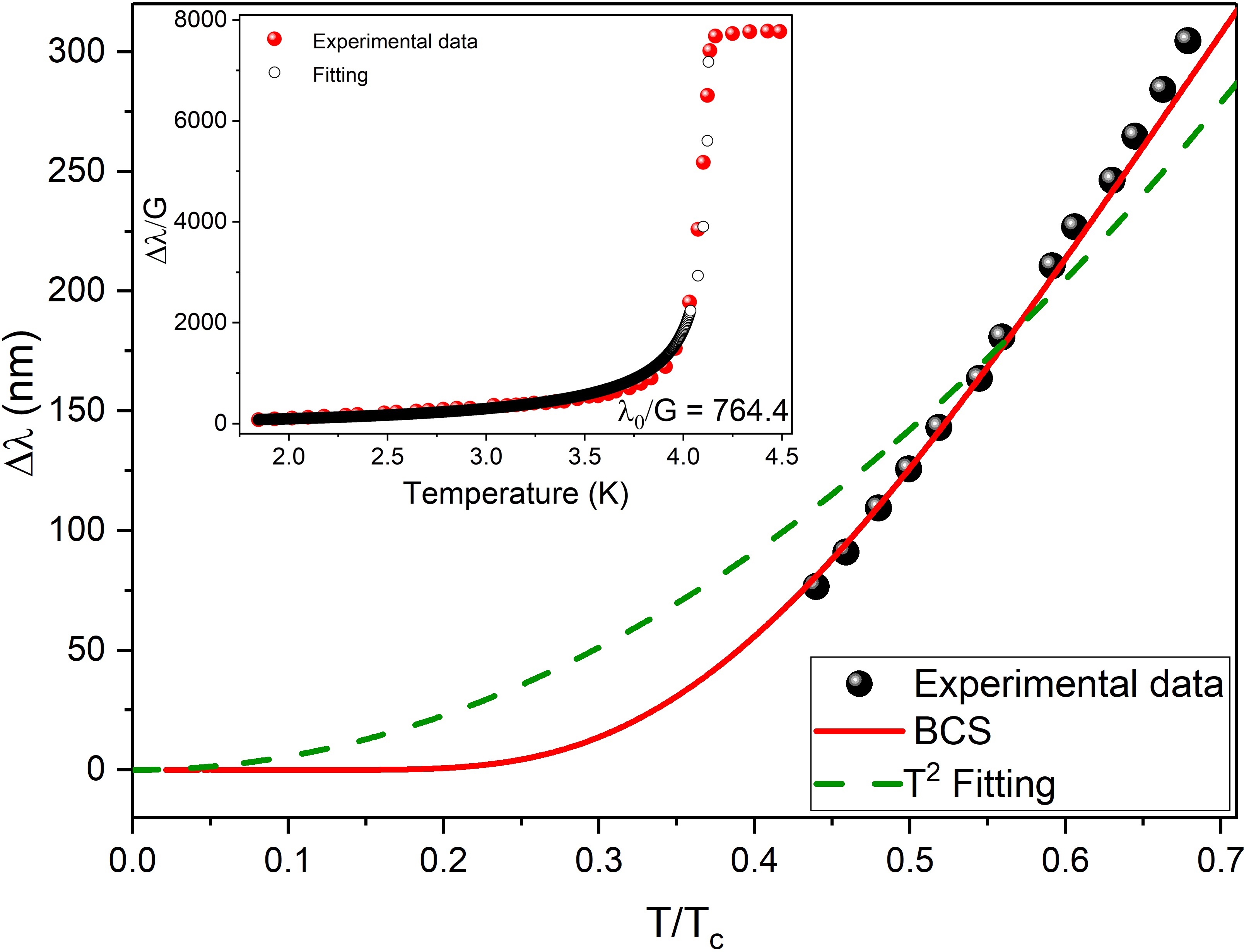} 
  \caption{Low temperature penetration depth dependence $\Delta\lambda$ as a function of reduced temperature. Solid red line shows the exponential fitting with fitting parameter as $\Delta_0/k_BT_c=1.85$ and dashed green line is $T^2$ fitting of the data. and inset shows Variation in rf penetration depth as a function of temperature to temperature in zero external field and fitting with classic two fluid model.}
\end{figure}
 \subsection{\label{app:subsec}Radio frequency penetration depth}
 The tunnel diode oscillator technique involves placing a superconducting sample inside a inductor which is a part of LC circuit driven to resonance by an oscillator operating at radio frequency. Compared to the lower critical field $H_{c1}$ of the sample, the amplitude of the rf field produced by the coil is significantly smaller (of the order of $\mu T$). Here with increase in temperature shift in resonance frequency of oscillator corresponds to increase in flux in the sample which can be directly related to the penetration depth. Fig. 5 shows change in penetration depth $\Delta\lambda$ that is equal to $G\times \Delta F(0, T)$, $G$ being the geometrical factor calibrated to be $2.27\si{Hz}$. Inset of Fig. 5 plots the variation in penetration depth as a function of reduced temperature. For an isotropic one-gap BCS model the $\Delta\lambda$ follows as;
 \begin{equation}
 \Delta\lambda(T)=\lambda(0)\sqrt{\frac{\pi\Delta_0}{2k_BT}}exp(-\frac{\Delta_0}{k_BT})
 \end{equation}
 where, $\Delta_0$ and $\lambda(0)$ are the values of the superconducting energy gap $\Delta$ and penetration depth $\lambda$ at absolute zero temperature respectively. This exponential relationship is largely relevant near $T_c$. For d-wave pairing in the clean limit, however
 \begin{equation}
 \Delta\lambda(T)\sim\lambda(0)\frac{2\ln2}{\alpha\Delta_0}T
 \end{equation}
 where $\alpha=\Delta_0^{-1}[d\Delta(\phi)/d\phi]_{\phi\to\phi_{node}}$ and $\Delta(\phi)$ is the angle dependent gap function. The d-wave gap is suppressed in the case of dirty limit, and the temperature dependency changes from linear to power law behaviour; $\Delta\lambda\sim T^2$.\\
  \begin{figure*}
\includegraphics[width=0.9\textwidth,height=12cm]{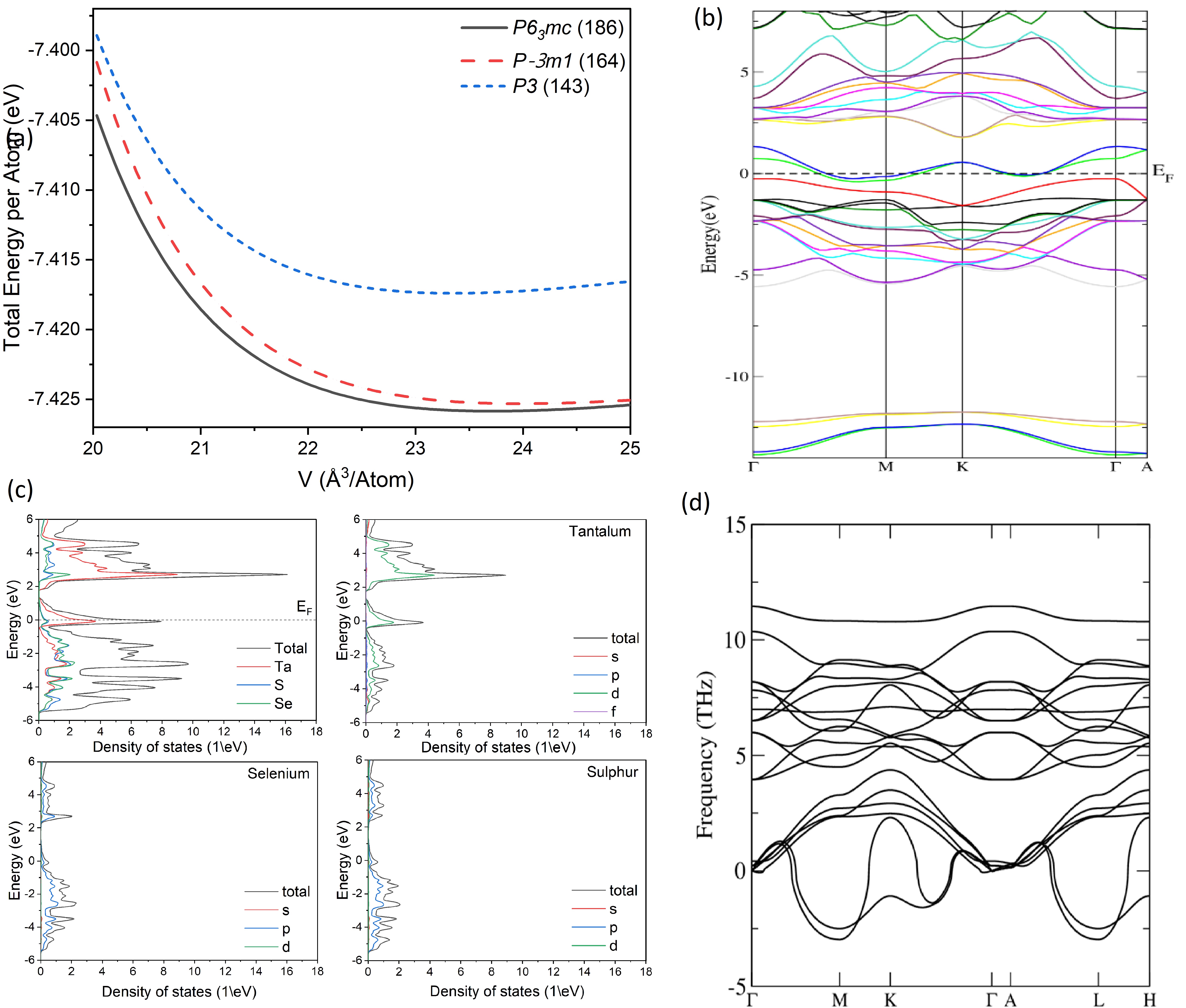} 
  \caption{(a) Equation of states curves for possible structures of TaSeS. (b) Electronic band structure. (c) Total density of states along with orbital contribution from all individual atoms. (d) Phonon dispersion curves.}
  \end{figure*}
 Fig. 5 displays low temperature dependency on penetration depth variation $\Delta\lambda$ of TaSeS together with the fittings of a standard BCS model (solid red line) and quadratic dependency on temperature (dashed green line). The BCS model offers better fit to the data. Further the gap ratio $\Delta_0/k_BT_c = 1.85\pm0.06$ and the associated energy gap value $\Delta_0\sim1 meV$. These values are rather close to the BCS value for weak coupling superconductor \cite{goyal2016single}.
  \subsection{\label{app:subsec}Density functional theory calculations}
  DFT with VASP and Wien2K codes were used to obtain equation of states curve, band structure, and density of states for TaSeS \cite{kresse1996efficiency,blaha2001introduction}. Fig. 6(a) shows the equation of states curves for three possible structure of the compound. With the stochiometric ratios of elements and lattice parameters of the crystals, the compound could belong to three space groups i.e., $P3$, $P\bar 3m1$ and $P6_3mc$. Structure optimization through equation of states fitting shows that structure $P6_3mc$ has lowest energy for a definite volume among all of three. This confirms the most stable structure of TaSeS belongs to $P6_3mc (186)$ space group and further DFT calculations were based on the same symmetry. Fig 6(b) shows the electronic band structure of 2H-TaSeS. TaSeS bulk DFT band structure features a few bands that cross the Fermi level. Thus, TaSeS is expected to be a superconductor and exhibits metallic behaviour in normal state similar to many TMDs. The overall density of states at the Fermi level is quite low, which might account for the low $T_c$ \cite{bezotosnyi2019electronic}. Also, in ideal crystal, the Fermi level is exactly midgap but even minimal amount of doping (i.e. impurities and other defects) can considerately shift the Fermi level towards the conduction or valence band. Further, only electron bands cross $E_F$ which shows the majority charge carriers are electrons. Fig. 6(c) shows the density of states (DOS) along with contribution from different orbitals. DOS confirms that Ta-5d orbitals predominate in the bands near Fermi level, $E_F$ as well as in conduction band.  Below $E_F$, a combination of several electronic states (Ta-5d, S-3p, and Se-4p) is seen, with a minor prevalence of Se-4p states. There is a crossover around $-0.5 eV$ where the S and Se DOS surpass the Ta DOS, showing that the S and Se states contribute more to the region near $E_F$ in valance band than the Ta states. The phonon dispersion spectra using a $3\times3\times2$ supercell of the bulk 2H-TaSeS are shown in Fig. 6(d). The negative values in the curves provide conclusive proof of structural instability in the compound.The CDW distortion as seen experimentally is thought to be directly connected to the phonon instability \cite{wang2020charge}.    
\section{\label{sec:level1}CONCLUSION}
    In brief, we report the growth of high-quality single crystals of TaSeS with chemical vapour transport method. The samples exhibit superconductivity with resistive and magnetic transition temperature of $4.15K$. The superconducting shielding fraction is about $98\%$ as calculated from ZFC susceptibility. There is evidence of charge density wave onset at $66K$. Both lower and upper critical fields fit well with Ginzburg-Landau equation. Calculated anisotropy factor is about $1.52$. Study of critical fields shows that TaSeS is a type-II superconductor with weak thermal vortex fluctuations. According to the Arrhenius law, the activation energy behaves in a way that is consistent with a power law in magnetic field, and the sample exhibits flux flow and flux pinning. At low temperatures, penetration depth $\Delta\lambda(T)$ shows exponential dependenc on temperature. The BCS exponent ratio, $\Delta_0/k_BT_c$ =$ 1.85$ which is slightly higher than the BCS value. DFT calculations with the help of VASP and Wien2K codes confirm the superconducting and normal state metallic behaviour of TaSeS. We found no evidence of topological superconductivity in TaSeS.\\
    \begin{acknowledgments}
    K Yadav and M Singh acknowledges Council for Scientific and Industrial Research (CSIR) for Junior Research Fellowship. M Lamba thanks University Grant Commission (UGC) for financial support through JRF. We express our gratitude towards Department of Science and Technology, Government of India for low temperature and high field facility at JNU (FIST program). We also thank Advanced Instrumentation Research Facility (AIRF), JNU for technical support.
\end{acknowledgments}

\bibliography{TaSeS}

\end{document}